# Performance Analysis of 6T and 9T SRAM

Ezeogu Chinonso Apollos
*Scholar, National Information Technology Development Agency, Nigeria.*

**Abstract**
The SRAM cell is made up of latch, which ensures that the cell data is preserved as long as power is turned on and refresh operation is not required for the SRAM cell. SRAM is widely used for on-chip cache memory in microprocessors, game software, computers, workstations, portable handheld devices due to high data speed, low power consumption, low voltage supply, no-refresh needed. Therefore, to build a reliable cache/memory, the individual cell (SRAM) must be designed to have high Static Noise Margin (SNM). In sub-threshold region, conventional 6T-cell SRAM experiences poor read and write ability, and reduction in the SNM at various fluctuation of the threshold voltage, supply voltage down scaling, and technology scaling in nano-meter ranges (180nm, 90nm, 45nm, 22nm, 16nm and 10nm). Thus, noise margin becomes worse during read and write operations compared to hold operation which the internal feedback operates independent of the access transistors. Due to these limitations of the conventional 6T SRAM cell, we have proposed a 9T SRAM that will drastically minimize these limitations; the extra three transistors added to the 6T topology will improve the read, hold and write SNM. The design and simulation results were carried out using Cadence Virtuoso to evaluate the performance of 6T and 9T SRAM cells.

## I. INTRODUCTION

Process-Variation-Aware SRAM architecture using the new 9T SRAM CMOS 45nm scaling technology node enables complete data isolation from the bit lines or memory cell thus preventing sneak path thereby providing more data read and write stability, reduced leakage power compared to 6T, 7T and 8T. We designed 6T and 9T SRAM cells to compare them in terms of stability and current leakage. The 9T configuration in this paper is a design paradigm for ultra-low power and robust logic circuit under process variation that reduces to the barest minimum dynamic and static current (power) consumption in 9T SRAM bit cell.

The paper will deliver the following using:
1. Peripheral components schematic and test bench of the simulation environment.
2. Result of read and write simulations of 6T SRAM and 9T SRAM.
3. The butterfly curve for both 6T and 9T SRAM with analysis of the derived Static Noise Margin (SNM), read and write N-Curves.
4. Power leakage test.

## II. LITERATURE REVIEW

### A. 6T SRAM

6T SRAM is the conventional SRAM design. This is made up of six transistors, whereby two of the transistors are PMOS type which then replace the resistive load used in 4T design. The configuration is such that the PMOS and NMOS form a cross-coupled inverter while two NMOS transistors are connected one each to the bit lines (see Figure 1a). Thus, these NMOS bit lines connected transistors are referred to as the "access transistors" which are controlled by the word line.

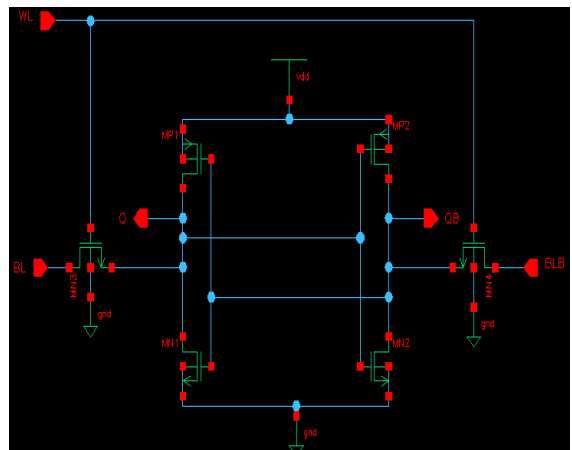

**Figure 1a: 6T Schematic Diagram [1].**

The SRAM has three basic operations: Read, Hold and Write operations.

*a). Read Operation:*

This is the state when data is requested from the memory cell. Thus, to read data, both bit line (BL) and bit line-bar (BLB) are initially pre-charged to a logic state 1 ($V_{dd}$), when the word line (WL=0) is low. After the pre-charge cycle the word line (WL) is enabled (WL=1) thus the access transistors (MN3 and MN4) are switched ON thereby connecting them to the bit lines [1]. Please note that the read operation is achieved by using the sense amplifiers that pull the desired data and produce the output; while the row decoders and column decoders select the appropriate cell or cells from which the data is to be read and are given to the sense amplifiers through transmission gate [1-2]. See figure 1b and 1c for simplified schematic during read 0 and 1 respectively.

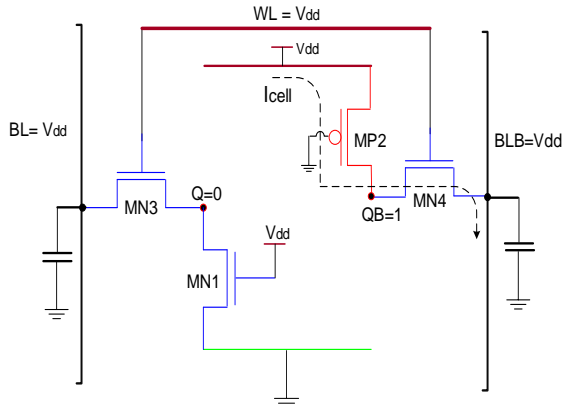

**Figure 1b: Read data path for Data=0 [1]**

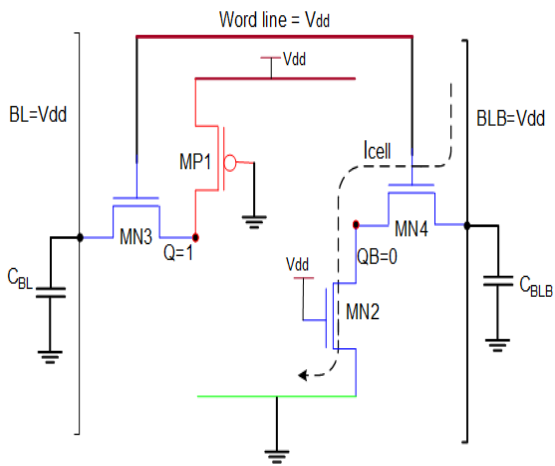

**Figure 1c: Read data path for Data=1 [1]**

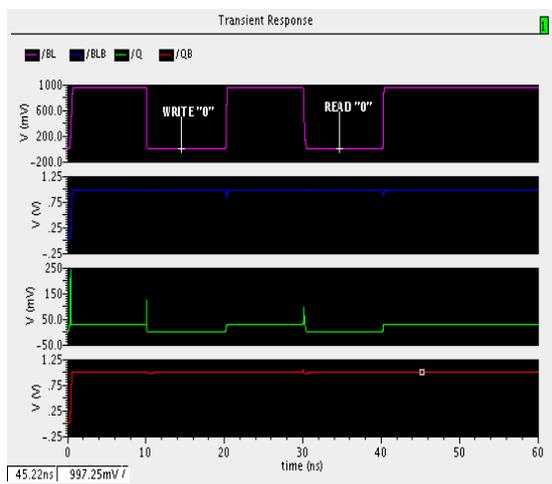

Figure 1d: 6T Write 0 and Read 0 [1]

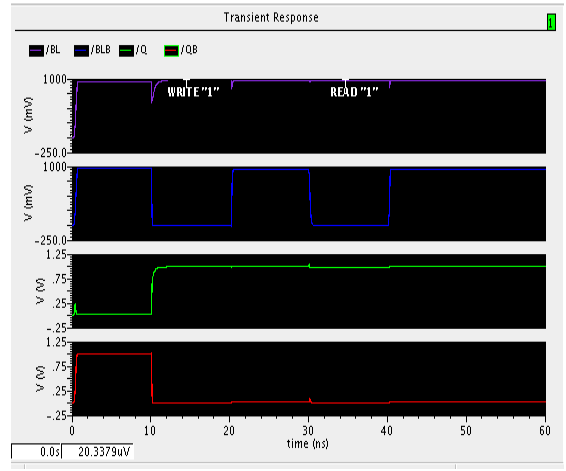

**Figure 1e: 6T Write 1 and Read [1]**

*b). Hold Operation*:

This is the state when the SRAM cell is idle (data is held in latch) and the bit line and bit line bar (data path) are kept at *gnd* when the access transistors are disconnected because the word line is not inserted. Thus, the PMOS transistors will continue to re-enforce each other as long as they are connected to the power supply in order to keep the data stored in the latch as shown in Figure 1f. Also from figure 1a, during this idle/retention mode, when "1" is stored in the cell, MP1 and MN1 are ON hence there exists a positive feedback between Q and QB nodes making Q to be pulled to $V_{dd}$. Similarly, when "0" is stored in the cell, MP1 and MN1 are OFF while QB is pulled to $V_{dd}$.

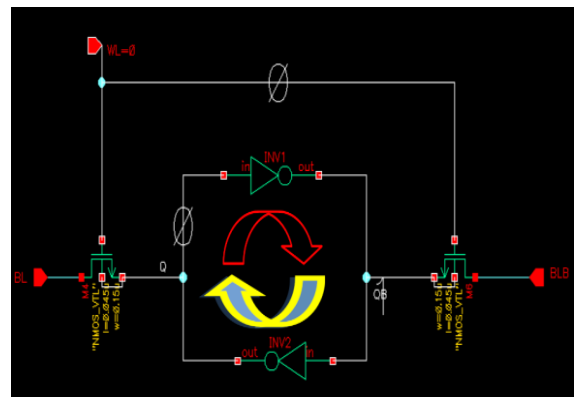

**Figure 1f: Retention Mode [1]**

*c). Write Operation:*

This is the state when data is written/updated in the cell (see Figure 1g). To write data into a cell, the sense amplifier and pre-charge circuits are deactivated while write enable and the word line are first activated then the input data is driven through the write driver input pin then the bit line is pulled to the value of the given data while the bit line bar (BLB) takes the complementary value. For instance, if data=0 then BL =0 while BLB = 1 ($V_{dd}$); whereas, if data=1 then BL =1($V_{dd}$) while BLB

= 0 (gnd). Hence, given that transistors MP1 and MN3 in figure 1g are correctly sized then cell will flip and the data is effectively written.

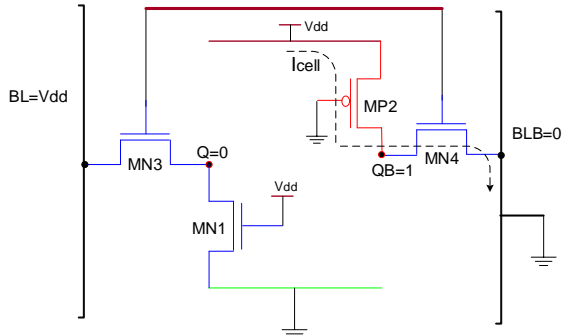

Figure 1g: Simplified Data Path for Write Operation (switching data 0 → 1)

### B. 9T SRAM

In this section, we present 9T SRAM proposed in [1,9] then carry out performance evaluation with the conventional 6T SRAM topologies in terms of stability, process variation and current leakage so as to justify the performance evaluation.

should be made symmetrical. During read operation, RWL is activated and transistors MN5 and MN6 (see figure 2a) are turned ON which will form strong pull down compared with conventional 6T SRAM. Thus, strong pull-down results to less resistance between data storage nodes to ground; therefore, the amount of raise in voltage of node Q will be less. The write operation is done by enabling the word line (WL) and disabling the read word line (RWL) then inserting the write enable signal.

During write operation, transistors MN5 and MN6 will allow only small leakage current because transistor MN7 will be cut-off, this is called SCCMOS technique. Furthermore, the SNM can be improved by increasing the width to length ratio of MN5 and MN6 [3].

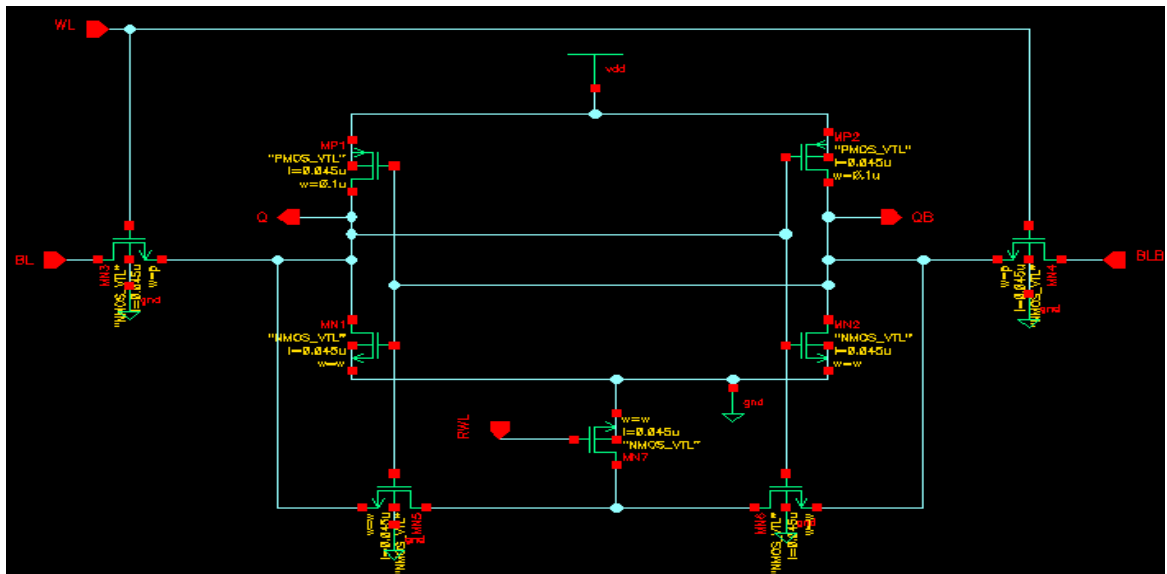

**Figure 2a: 9T Schematic Diagram [1].**

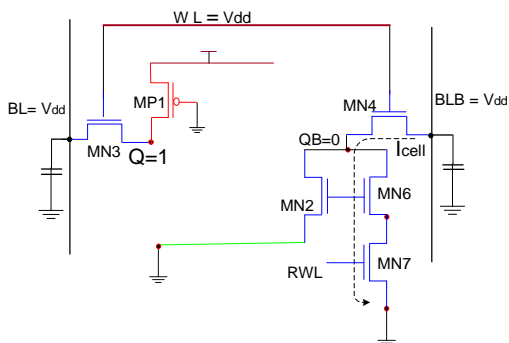

Figure 2b: 9T Read path for Data=1

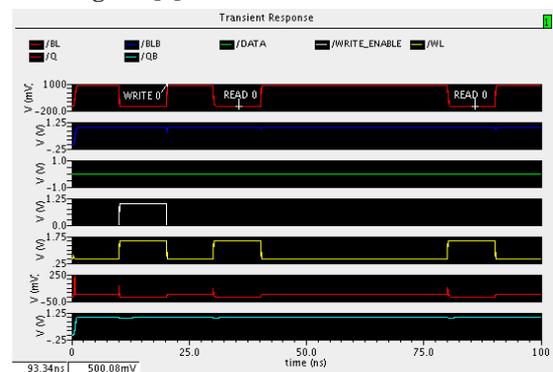

**Figure 2c: 9T Write Data =0, Read Data =0**

This configuration employs a differential read operation for better read access time and the design

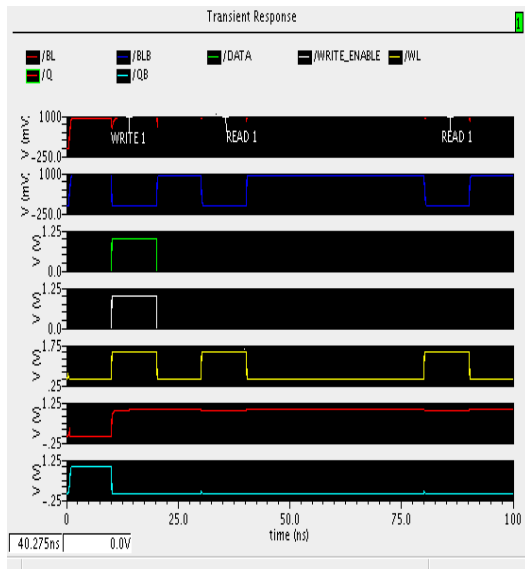

**Figure 2d: 9T Write Data =1, Read Data =1**

## III. PERFORMANCE CRITERIA

*A. Stability Metrics*

This paper explored the use of SNM metric and N-curve metric for the stability analyses and evaluation of the 6T and 9T SRAMs design.

*a) Static Noise Margin (SNM)*:

The stability of SRAM cell is mainly defined by the use of SNM which is the maximum value of DC noise voltage that can be tolerated without changing/flipping the internal storage state of the SRAM. In this paper the graphical approach was used and the value of the SNM will be the length of side of the largest embedded square on the butterfly (VTC) curve. The schematic setup for the SNM simulation is given in figure 3a and 3b for 6T and 9T respectively which is known as the "worst case SNM setup".

*1). Hold Static Noise Margin*

To determine the Hold SNM, the following steps were taking:
1. Connecting BL and BLB to $V_{dd}$;
2. Connecting WL to gnd;
3. Plotting both VTCs of the inverter 1 and inverter 2 on the same graph.
4. Finding the maximum square SRAM that can fit into the VTC lobe.
5. The SNM is the side of the maximum square.

The hold SNMs for the 6T and 9T SRAM are the same because the same symmetry of 6T SRAM design was used with additional three extra transistors to improve the read margin in the 9T SRAM design. These transistors are cut-off during hold state, therefore given an equivalent 6T SRAM configuration (see figure 4a and 4d for hold SNM).

*2). Read Static Noise Margin (RSNM)*

1. Connecting BL and BLB to $V_{dd}$;
2. Connecting WL to $V_{dd}$;
3. Using graphical method, the SNM is the side of the maximum square (see figure 4b and 4c).

*3). Write Static Noise Margin (WSNM)*

1. Connecting BL to gnd and BLB to $V_{dd}$;
2. Connecting WL to $V_{dd}$;
3. Using graphical method, the SNM is the side of the maximum square.

The Write operation of the proposed 9T SRAM is equivalent to the 6T SRAM operation because the RWL is deactivated which thus cut-off transistors MN6 and MN7 therefore making the configuration a 6T SRAM [1]. See figure 4d, 4g and 4h for the simulation result.

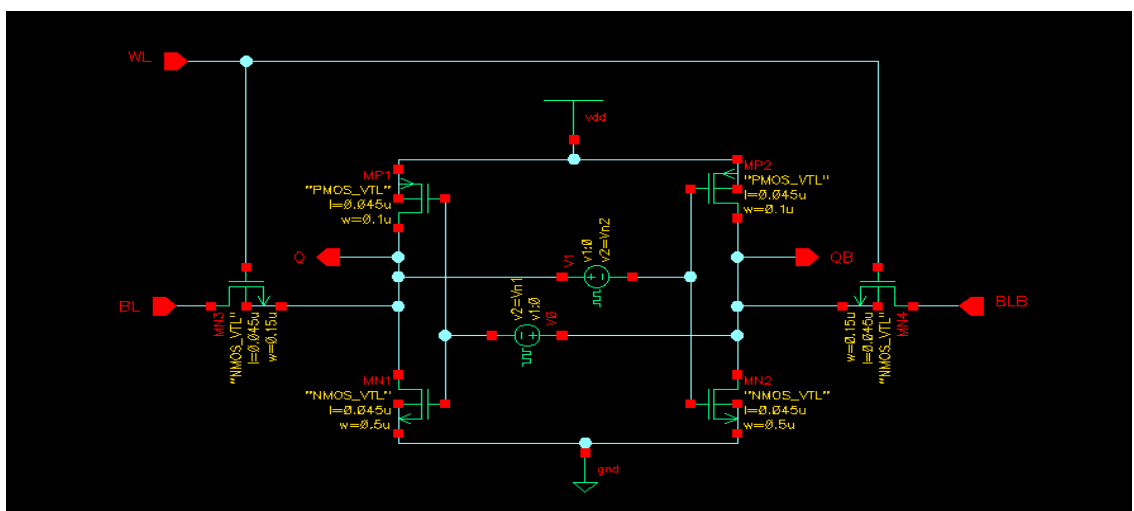

**Figure 3a: 6T Stability Noise Margin Setup** [1].

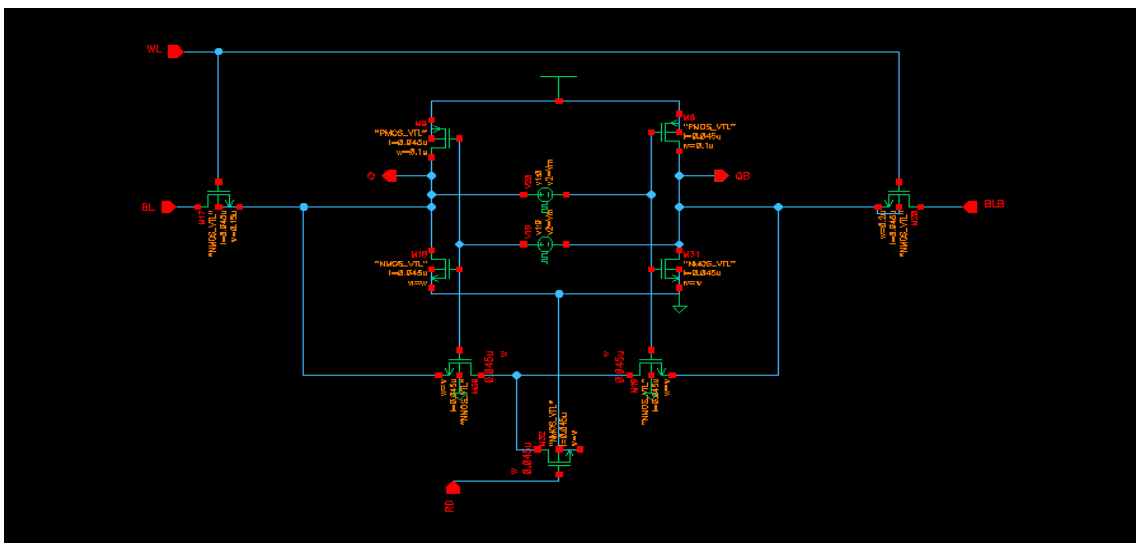

**Figure 3b: 9T Stability Noise Margin Setup [1].**

*b). N-Curve Metric:*

The cell stability depends on supply voltage, therefore as supply voltage scales down the cell becomes less stable. The SNM technique has been in use for quite some time as a metric for measuring stability of SRAM cell by drawing butterfly (or VTC) curves of the two back-to back inverters of the SRAM from a DC simulation. However, the disadvantage of measuring the SNM using butterfly curves (VTC) approach is the inability to measure the SNM with automatic inline testers and also more time consuming due to the mathematical calculations or the fitting of the squares on each lobe of the VTC curve to determine the SNM. And sometimes it may not give very accurate result due to systematic errors in computation. In addition, it is quite rigorous and time consuming. Whereas N-curve metric is used for inline testers; it gives both information for voltage and current [5] and in addition it has no voltage scaling delimiter as found in VTC approach. It also has all the information about the SRAM stability and write ability in a single plot. In addition, N-curve can be extended to power metrics both the voltage and current information are taken into account to provide better stability analysis of an SRAM cell [4].

There are many factors that affect the stability of an SRAM, these include: Pull-up Ratio, Cell Ratio, Supply Voltage, Temperature, Technology variation:

• *Pull-up Ratio*: The write margin directly depends on the pull up ratio [6]. As pull up ratio increases, WSNM gets reduced, therefore pull up ratio should not be increased beyond certain limit [7]. For better WSNM the access transistors should be stronger than the pull up device (PMOS).

• *Cell Ratio*: The read margin depends more on the cell ratio. Therefore, the larger the cell ratio the higher the SNM. Thus, strong driver transistors and weak access transistor is preferable for better SNM. See Figures 4a, 4b and 4c in section IV for variation of SNM with Cell Ratio simulation result.

• *Supply Voltage*: Read operation becomes destructive at very low voltages. The read speed, read and write margin are reduced to a great extent when the supply voltage is scaled down close to the $V_{th}$.

• *Temperature*: SRAM cell's speed increases with temperature increase but the SNM reduces [10].

• *Technology Variation:* Device dimension is reducing, thus leading to fluctuation of intrinsic process parameters (random dopant density variation in channel, drain and source) which results in variation of $V_{th}$ which affects SRAM cell stability to a large extent and also write time [7] and hence also reduces with technology scaling reduction to 45nm, 22nm and 16nm nodes.

Meanwhile, figure 3c and 3d are the setup for the N-curve for 6T and 9T SRAM designs, however it can also be used to determine the SNM - when one DC noise source is used then it is called the "best case SNM". A "dc sweep" was then run to determine the HSNM, RSNM and WSNM. Note that figure 3c shows the setup of 6T SRAM using the memory cell; while figure 3d is for the 9T SRAM cell using another set up approach after converting the 9T SRAM memory cell to its equivalent block diagram. The 6T SRAM could be done likewise by the researcher.

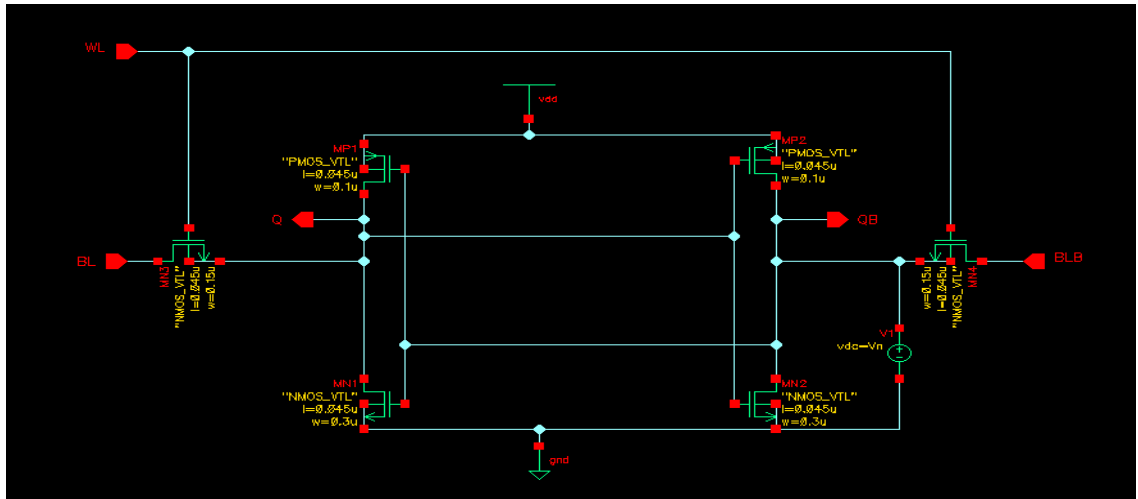

**Figure 3c: 6T SRAM N-Curve Simulation Setup [1].**

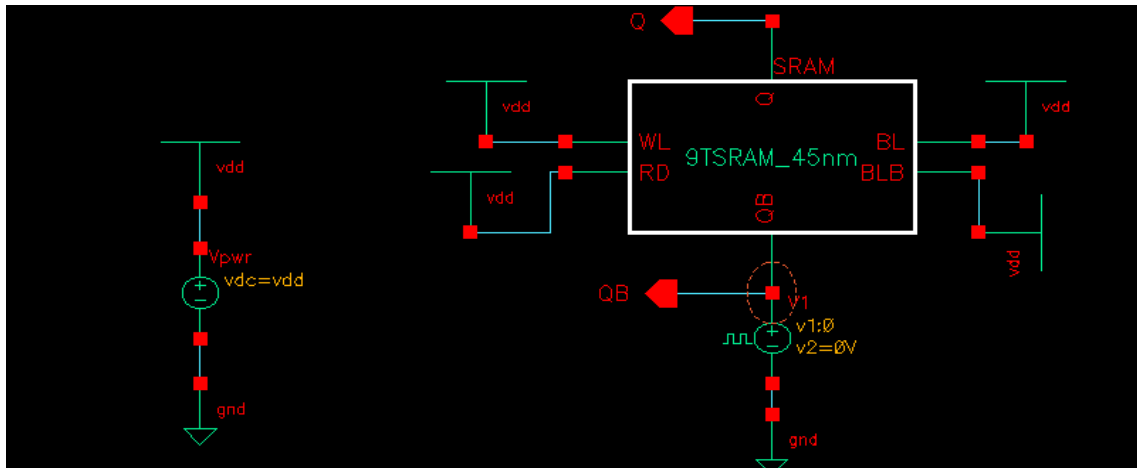

**Figure 3d: 9T SRAM N-Curve Equivalent Simulation Setup [1].**

*1). Read Stability*

The read stability was found by extracting the N-curve during read operation. To do this, the bit-lines are clamped at $V_{dd}$ and the word-line activated to put the SRAM in the read mode. Then, a voltage sweep of $V_{in}$ from 0V to $V_{dd}$ was applied at the node QB assume it is storing a "0" to obtain the corresponding current $I_{in}$. The simulation results for the N-curve is presented here for 6T and 9T SRAM cells in figure 5a and 5b. To get the read N-curve, the test bench connection as shown in figure 3c and 3d are used.

*2). Write Ability*

The SRAM write ability can be determined using the write N-curve. This is similar to the read N-curve for WTI except that one of the bit lines, that is BLB or BL depending on where the probe for dc sweep is attached, is clamped to gnd instead of $V_{dd}$ as for read N-curve [1]. Thereafter, dc sweep is performed on the internal node QB (in our setup instance); the write N-curve helps us to get the critical current ($I_{CRIT\_WR}$) which is the minimum current for write operation. In other words, the $I_{CRIT\_WR}$ derived from the curve is the critical current required to write data into the cell without failure.

*3). Static Voltage Noise Margin (SNVM)*

It is the maximum tolerable DC noise voltage at the input of the inverter of the cell prior a change in its content. Thus, from the simulation plot, the voltage difference between A and B in figure 5a shows maximum tolerable DC noise voltage before flipping of content of cell [1].

$$SVNM = V_B - V_A \quad \ldots\ldots\ldots\ldots\ldots\ldots (1)$$

*4). Static Current Noise Margin (SIVM)*

It is the maximum tolerable DC current that can be injected into the cell prior a change in its content; and it is measured as a peak current located between point A and B. The SINM is used to characterize the cell stability [1,3].

*5). Write Trip Voltage (WTV)*

This is the minimum voltage drop needed to flip the internal node "1" of the SRAM cell when both bit lines are clamped to $V_{dd}$. Consequently, it can be measured as a difference between point C and B as shown in figure 5a and 5b simulation result. WTV is used to characterize the write ability of the cell [1,3].

*6). Write Trip Current (WTI)*

It is the minimum amount of current needed to write the cell and can be measured as a negative peak current between C and B as shown in the N-curve of figure 5a and 5b simulation result. An overlap of points A and B or point B and C means loss of stability of SRAM cell and WTI is used to characterize the write ability of the cell [1,3].

*7). Static Power Noise Margin (SPNM)*

This is the product of the SVNM and SINM. And this value should be large for better write ability [3]. It is derived from the area below the curve between point A and B. And it can be described by the equation,

$$SPNM = \sum_{A}^{B} I_{in} * V_{in} \quad \ldots\ldots\ldots\ldots (2)$$

- *Write Trip Power(WTP)*

This value should be large for better write ability [3]. It is derived from the area below the curve between point C and B. Also, it can be described by the equation,

$$WTP = \sum_{B}^{C} I_{in} * V_{in} \quad \ldots\ldots\ldots\ldots (3)$$

where $V_{in}$ is the sweep voltage source and $I_{in}$ is the current supplied by the sweep voltage, $V_{in}$. Therefore, for successful read and write SPNM must be positive while WTP must be negative [3].

*B. Current Leakage Analysis*

As CMOS technology keep scaling down, leakage current is becoming a major source to the total power consumption of SRAM cells. Thus, in nano-scaled transistors with low threshold voltages, subthreshold and gate leakage are now the most dominant sources of leakage and keeps increasing as technology scales down to 22nm, 16nm and 10nm nodes. Threshold voltage of MOSFET reduces as the channel length is decreased. This reduction of threshold voltage with reduction of channel length is known as roll-off [8]. And the consequence is larger subthreshold current. In addition, gate leakage due to variation in oxide thickness, $t_{ox}$, and gate length and this has a severe effect on NMOS than PMOS devices due to exponential dependence on $t_{ox}$ [9].

*a) Current Leakage Reduction Schemes:*

In an SRAM design current leakage has majorly two dominant paths, $V_{dd}$ to gnd and bit line to gnd for a six-transistor SRAM cell. Leakage through these two paths consist of a high percentage of the total leakage [11]. Therefore, in order to reduce current leakage, various schemes have been proposed as follows:

• *MSRAM:* Memristor based SRAM which consist of 9T SRAM and 2T memristor switch and an isolation transistor using a dual read assist pass gate scheme [1].

• *MTCMOS:* Multi-threshold CMOS (MTCMOS) uses high $V_{th}$ as a cut-off MOSFET arranged in series connection with low- logic circuits to cut-off leakage current in a stand-by mode [4]. Unfortunately, MTCMOS does not work below 0.6V supply voltage due to the high- MOSFET threshold voltage, $V_{th}$, it does not turn ON; so it cannot be used for low voltage design of 1V.

• *VTCMOS*: Applies back-gate bias to cut off leakage current during stand-by mode by using the body effect [12]. "This scheme cannot be applied to fully depleted SOI process technology. It is also difficult to apply to partially depleted SOI process technology due to the overhead required to connect the body of each MOSFET with interconnection for applying the body bias" [13]. The second drawback of this scheme is that the VTCMOS requires modification to cell libraries to separate back-gate bias lines from and lines [13].

• *DTMOS*: Dynamic-threshold MOS (DTMOS) scheme ties the gate and body of a SOI MOSFET together, this makes the Vth of the MOSFET high in an off-state and low in an ON-state. The DTMOS, however, suffers from 10-mA-order leakage current in 0.5–0.7-V for 1 million logic gate VLSIs, due to inherent forward-bias current of the p-n-junction associated with the source–body junction of the MOSFET. By combining the SCCMOS and the DTMOS, the leakage current in a stand-by mode can be reduced while the DTMOS remains at high speed in an active mode. Because of this, the VTCMOS cannot be used with the DTMOS that have the body always tied to the gate [13].

• *SCCMOS*: The SCCMOS was proposed by [13] to be realized in CMOS logic circuits working below 0.5-V while maintaining 1-pA-order stand-by current per logic gate. The SCCMOS can be effectively combined with SOI, DTMOS, and/or PTL gates [13]. It uses low transistors with an inserted gate bias generator. The NMOS or PMOS insertion scheme can be used this is called "sleep control scheme" using either or both a PMOS or an NMOS connected to the power supply and gnd respectively.

• Data Retention Gated-ground cache (DRG): This puts the portion of the SRAM core to low leakage mode to reduce power. This is achieved by adding an extra NMOS transistor in the leakage path from the supply voltage to the ground of the static random access memory (SRAM) cells. This transistor turns on in the used sections while off in the unused sections, essentially "gating" the supply voltage of the cells [13]. This reduced leakage is due to the self-reverse biasing of the stacked transistors formed between the SRAM NMOS transistors and the Gated ground control transistor. DRG Cache technique utilizes inherent idleness of cache/SRAM to save leakage by turning off the idle sections of the SUM core and data is not lost when the gated- Ground transistor is turned off in the unused sections of the cache/SRAM [15].

Furthermore, other methods of reducing leakage are at the process technology level, well-engineering techniques by retrograde and halo doping are used to reduce leakage and improve short-channel characteristics. Consequently, at the circuit level, transistor stacking techniques, multiple and dynamic $V_{th}$, multiple and dynamic $V_{dd}$ techniques are used to effectively minimize the leakage current in high-performance logic and memory designs [13].

*b). Finding Current Leakage Using Cadence Tool*

In order to find the leakage current of an SRAM, a block symbol is created for the SRAM schematic then an NMOS transistor source is connected to the leakage node. Thereafter a DC voltage source of 0V is applied to the gate of the NMOS transistor while every other nodes of the circuit are connected to gnd [14]. See figure 6a and 6b. Thus, to run the simulation, the following steps
were employed [1]:
• *Choosing the dc analysis then checked "Save DC Operating point"; then OK.*
• *Run simulation.*
• *From the simulation window: selecting*
*Tools→Results Browser.*
• *Clicking on dcOpInfo-Info→V1 to see the information about the leakage current.*

*Alternatively, a transient simulation could be run, then the wave form will indicate the leakage current magnitude.*

*C. Process Variation*

Process variations are the critical design parameters – die to die and intra-die variation – from equipment processing in the semiconductor design technology due to inability to precisely control the fabrication process at small feature technologies at the nano-scale which in turn results in large variation in the operation and functionality of the design. This is very severe in the case of memory components as minimum sized transistors are used in their design [16]. These variations include the film thickness, lateral dimensions, supply voltage, doping concentration and threshold voltage variation. All these contribute to the circuit optimization for performance and power consumption. Doping concentration affects the threshold voltage, the $V_{th}$ increases steadily as a result of more random dopant fluctuations in channel, source and drain due to increase delay distribution and delay spread. Consequently, these random and systematic fluctuations affect the stability of the SRAM [1].

Thus, in the 6T SRAM design, the read stability of the cell is determined by the ratio of the current produced by the access transistors MN3 and MN4 (see figure 1a or 1b). Furthermore, the impact of variation increases as the supply voltage, $V_{dd}$, scales down to $V_{th}$ because the sensitivity of the circuit delay amplifies. Temperature and voltage variation are environmental variations which are primarily a function of intra-die (within die) variations, and contribute to failure rate (write ability and read stability) in SRAM cells.

## Iv. SIMULATION RESULTS

This section of the paper presents the output curves of the simulation of the stability measurement using Static Noise Margin and N-curve techniques under process variation such as temperature, cell ratio, pull-up ratio, and voltage supply for both the 6T and 9T Static Random Access Memory cells. Furthermore, the leakage current and power analysis setup and simulation results are shown and this can be seen from the setup and result of figure 6a and 6b for 6T and 9T SRAM cells respectively. The simulation is carried out under the same parameter testing conditions for both the 6T and 9T SRAM cells to ensure accuracy and uniformity in the performance analysis for both designs. The findings are presented as follows:

*A.   Static Noise Margin Simulation Results*

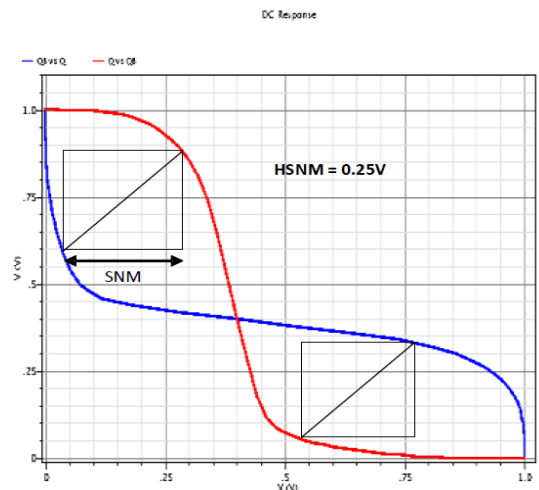

**Figure 4a. 6T and 9T Hold SNM for CR=1.5**

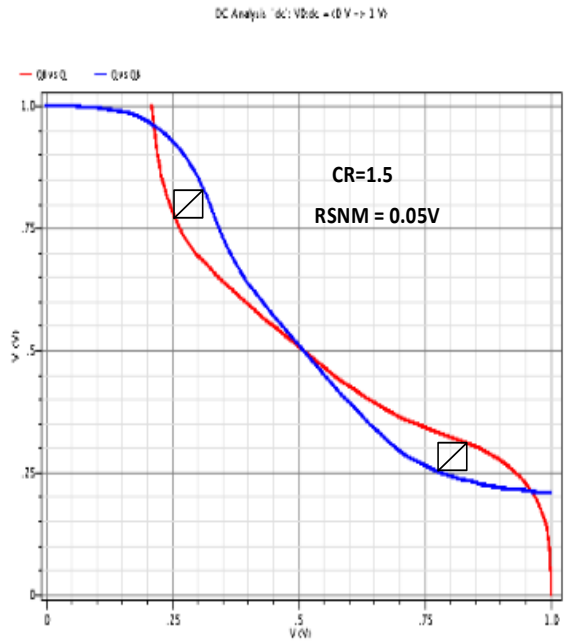

Figure 4b. 6T RSNM for CR=1.5

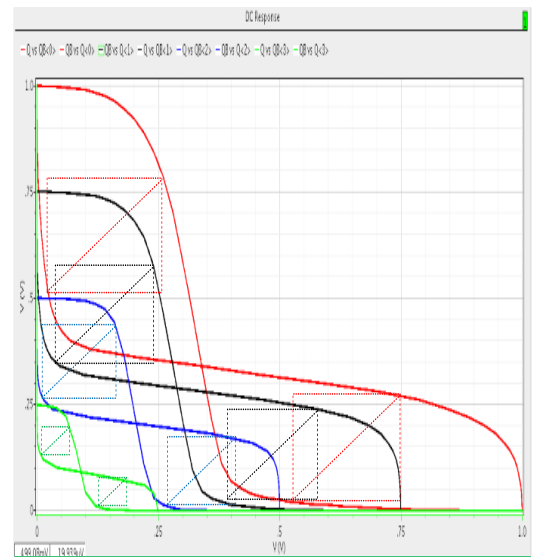

Figure 4d. 6T and 9T HSNM under Voltage Variation

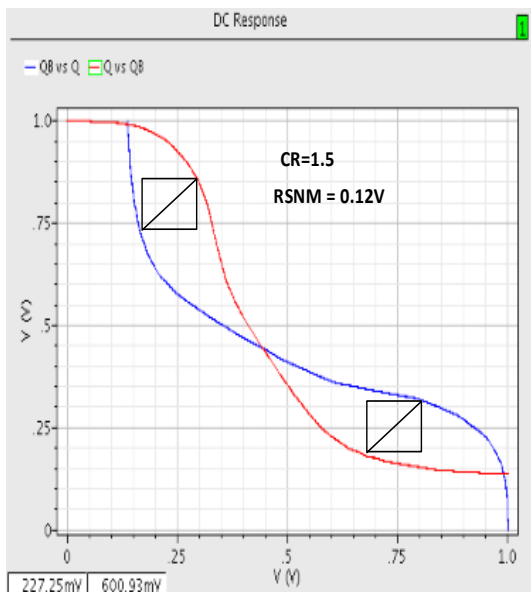

Figure 4c. 9T RSNM for CR=1.5

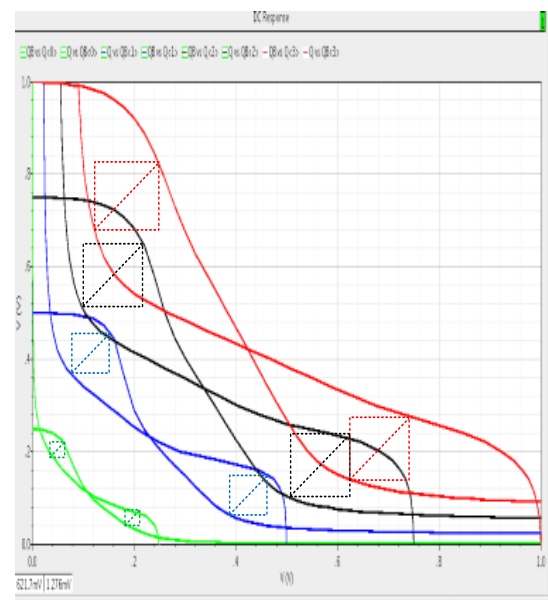

Figure 4e. 6T RSNM with Voltage Variation

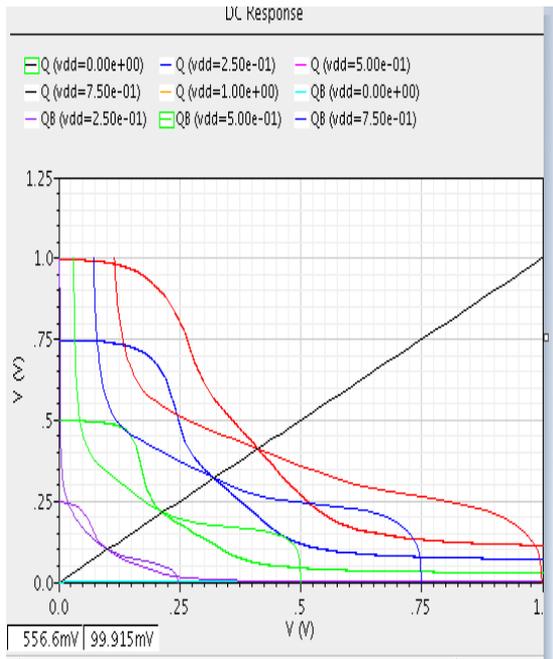

**Figure 4f. 9T RSNM with Voltage Variation**

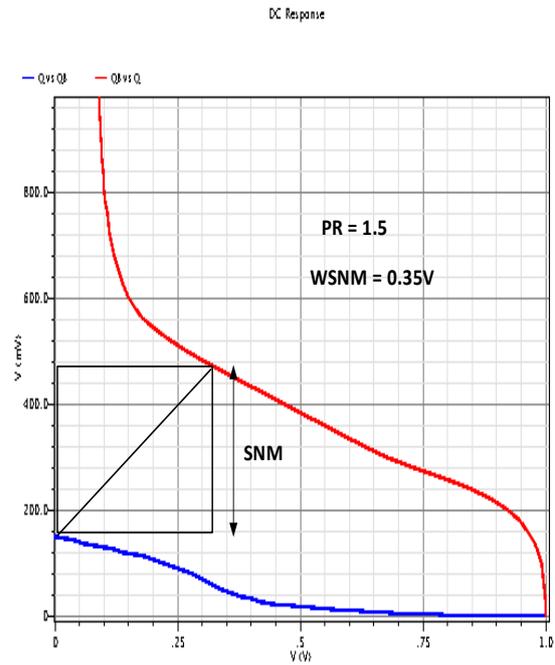

**Figure 4h. Write 1 WSNM for PR=1.5**

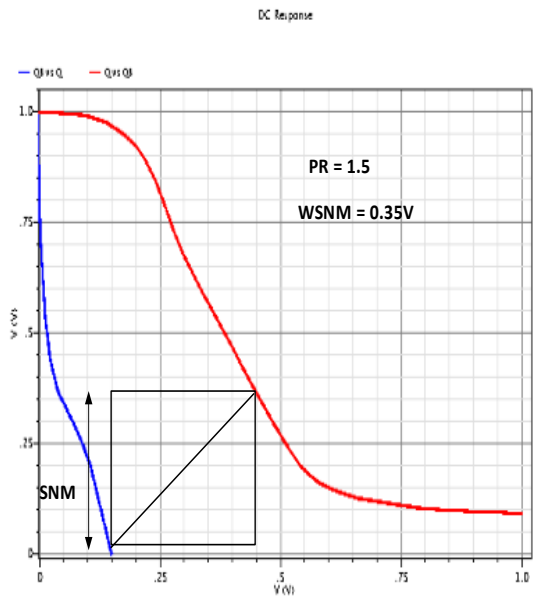

**Figure 4g. Write 0 WSNM for PR=1.5**

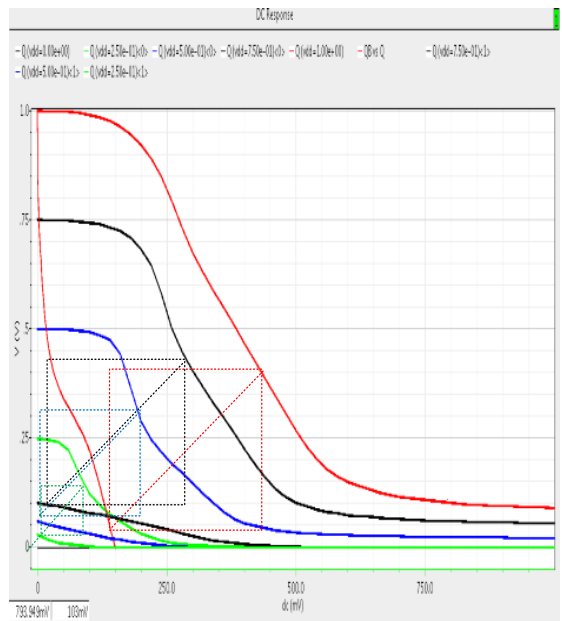

**Figure 4i. WSNM with Voltage Variation**

*B.  N-curve Simulation Results*

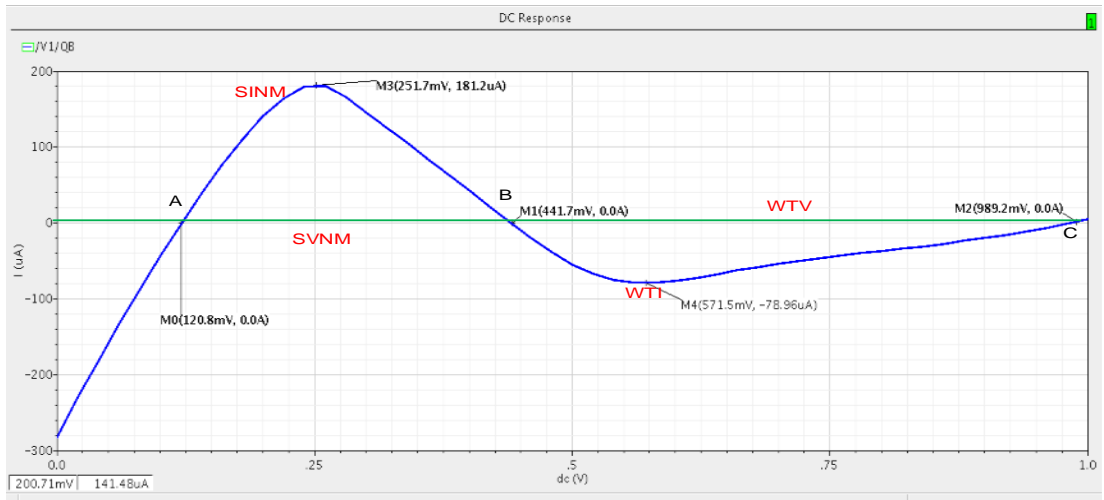

**Figure 5a: 6T Read N-Curve Simulation [1].**

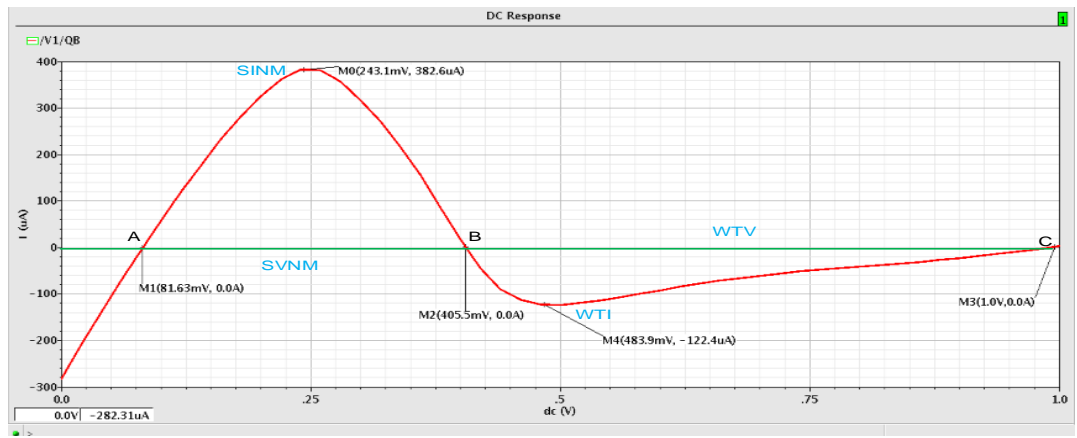

**Figure 5b: 9T Read N-Curve Simulation [1].**

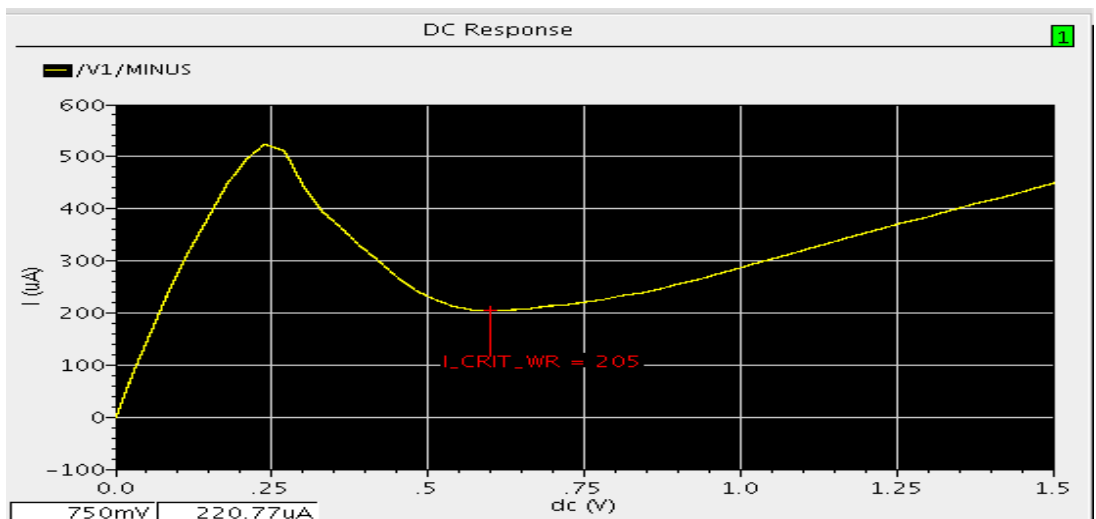

**Figure 5c: 6T and 9T Write N-Curve Simulation [1].**

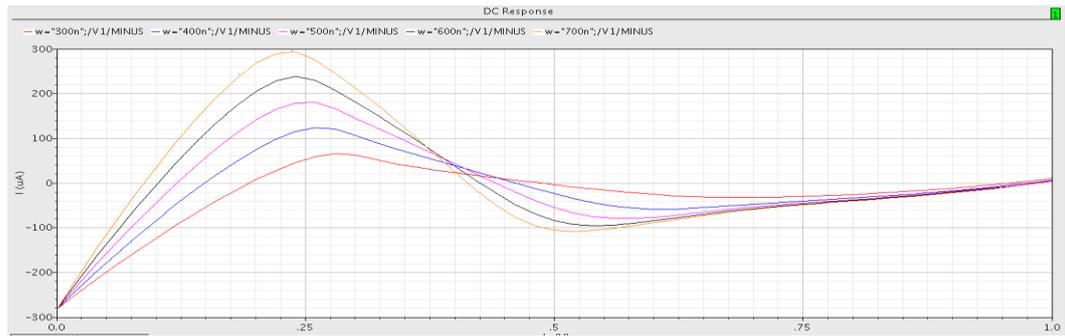

**Figure 5d: 6T SRAM Stability Variation with Cell Ratio [1].**

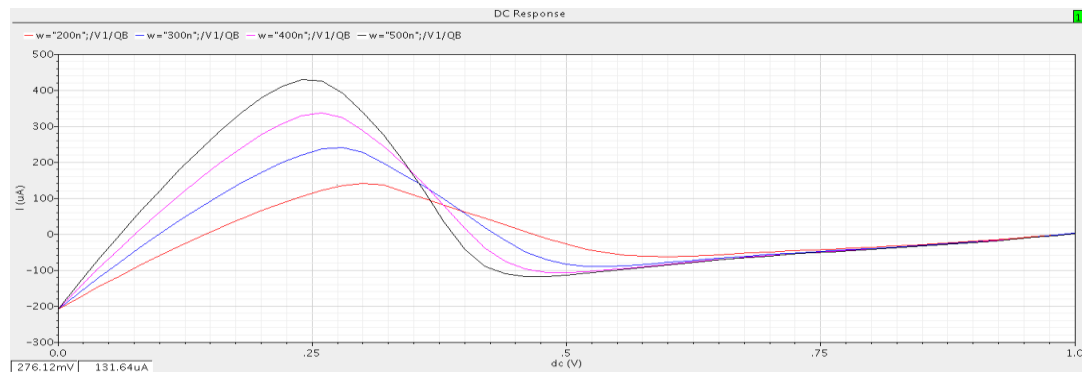

**Figure 5e: 9T SRAM Stability Variation with Cell Ratio [1].**

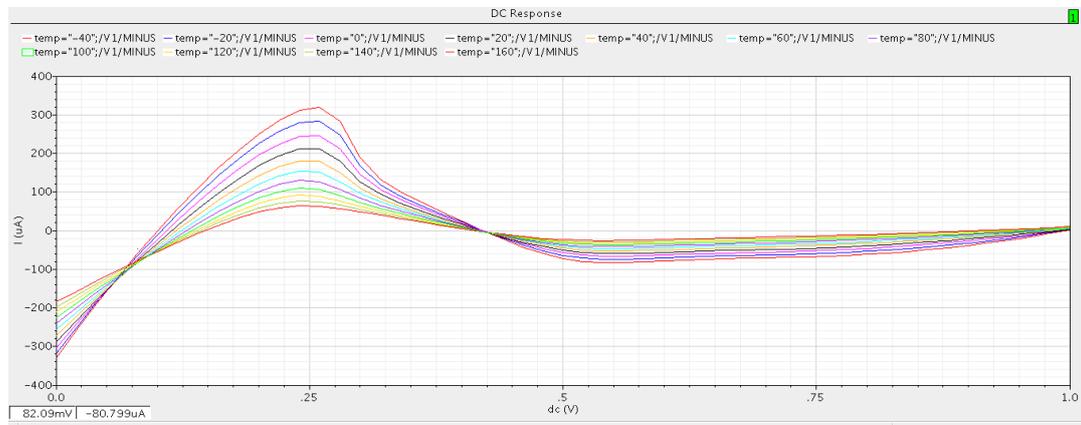

**Figure 5f: 6T SRAM Stability Variation with Temperature [1].**

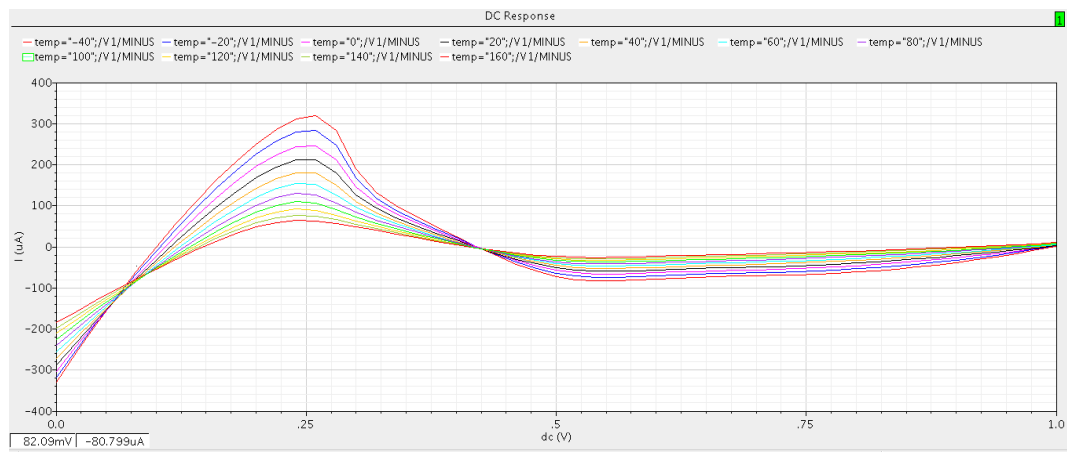

**Figure 5g: 9T SRAM Stability Variation with Temperature [1].**

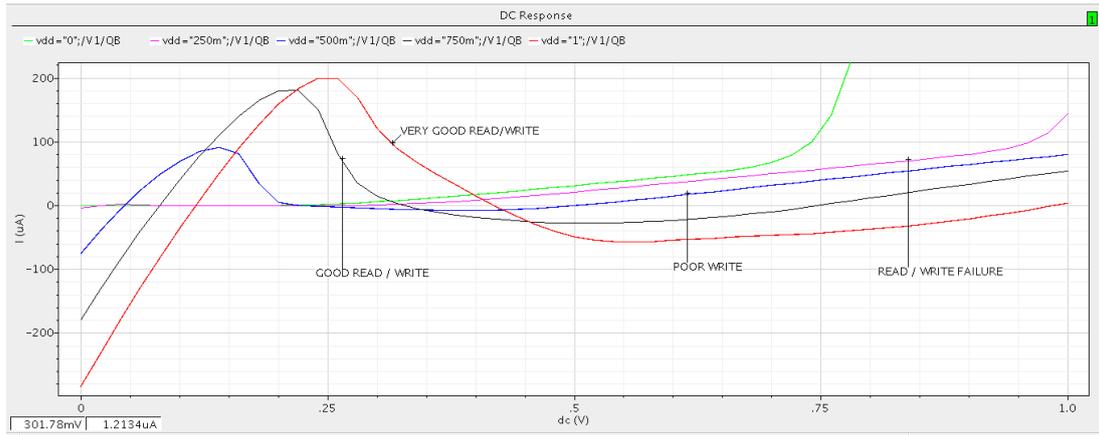

**Figure 5h: 6T SRAM Stability Variation with Supply Voltage [1].**

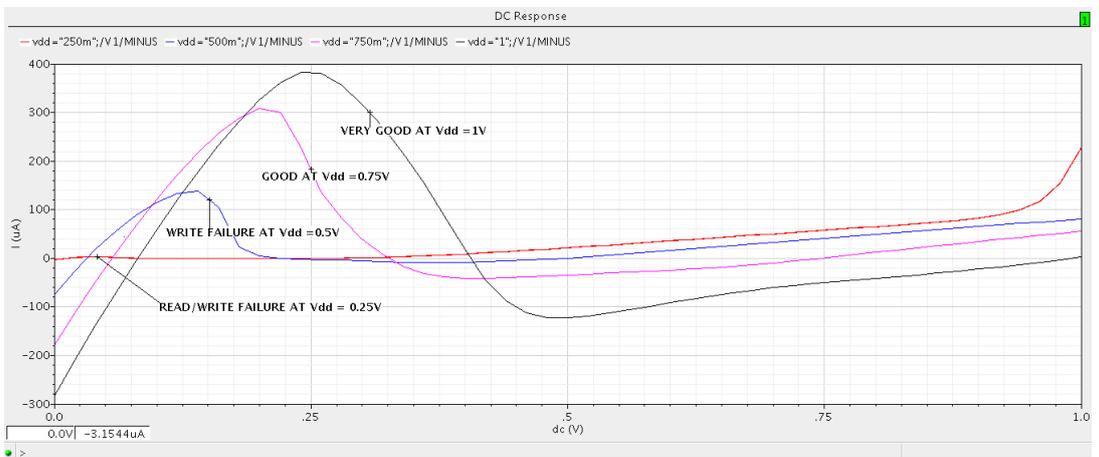

**Figure 5i: 9T SRAM Stability Variation with Supply Voltage [1].**

*C. Current Leakage Simulation*

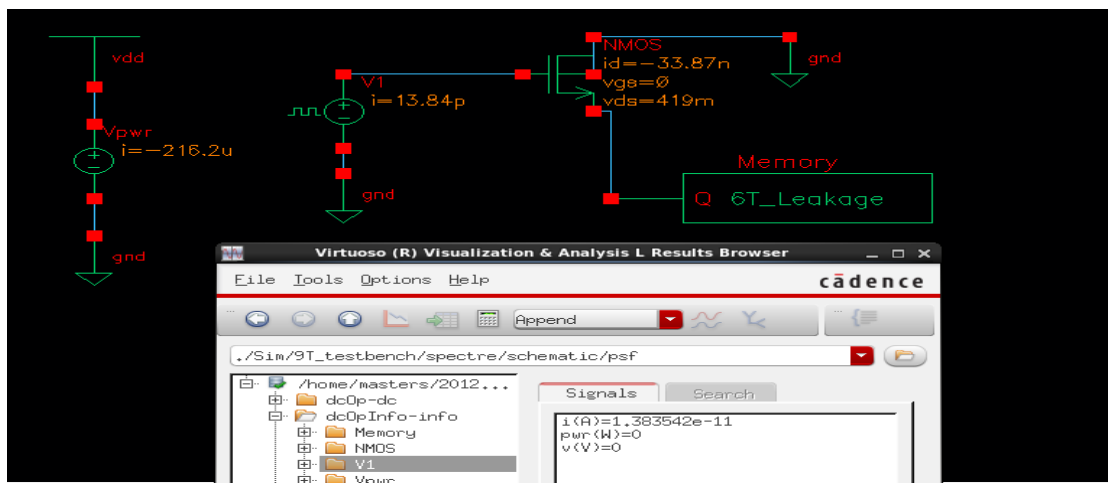

**Figure 6a: 6T Current Leakage Setup Simulation [1].**

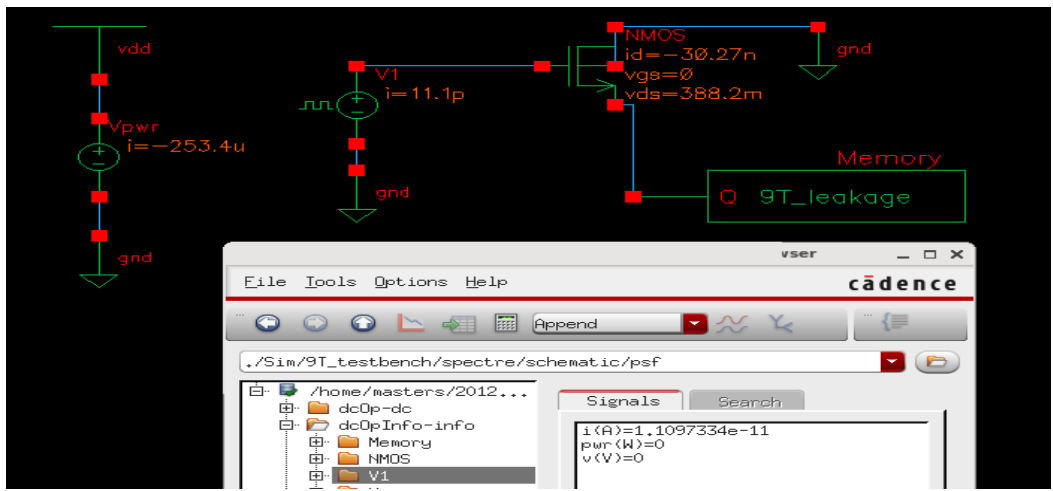

**Figure 6b: 9T Leakage Current Simulation Setup [1].**

## V. SUMMARY OF SIMULATION RESULT

In this paper the 6T and 9T SRAM have been compared using different performance criteria such as stability using both signal noise margin and N-curve techniques; process variation, variation with voltage and temperature; and finally the leakage current or power in the SRAM cells. Figure 7 indicates different cell ratio plot against Static Noise Margin (SNM); while table 1 gives the performance analysis summary of the research between the conventional 6T SRAM and the chosen 9T SRAM configuration.

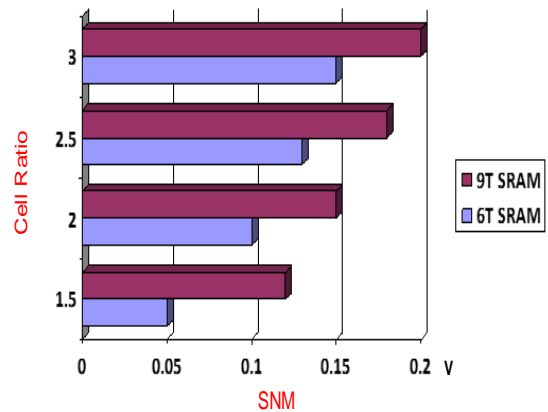

**Figure 7: 9T Leakage Current Simulation Setup [1].**

**Table 1: Summary of N-curve Stability Evaluation of 6T and 9T SRAM cells [1].**

| Parameters | Conventional 6T SRAM | Proposed 9T SRAM |
|---|---|---|
| CMOS Process | 45nm/1V, 27°C | 45nm/1V, 27°C |
| Read/Write Process | Differential | Differential |
| Stability Metric | N-Curve | N-Curve |
| SVNM | 320.9mV | 321.67mV |
| SINM | 181.2 $\mu A$ | 382.6 $\mu A$ |
| WTV | 547.5 mV | 594.5 mV |
| WTI | −78.96 $\mu A$ | −122.4 $\mu A$ |
| $I_{CRIT\_WR}$ | 205 $\mu A$ | 205 $\mu A$ |
| Read Leakage Current | 13.84 pA | 11.1pA |
| Non-Volatile State | No | No |

## VI. CONCLUSIONS

This paper presents 6T and 9T SRAM memory designs in 45nm CMOS technology node. The performance evaluation of both designs was carried out. The performance criteria were stability, power or current leakage and process, voltage and temperature variation.  And also, Static Noise Margin (SNM) and the N-curve metric were used for stability criteria; however, the N-curve was more preferred for the stability measurement because of the advantages of information the N-curve provides about voltage, current, and power in a single plot. The 9T SRAM was found to have a higher stability in the read margin this is due to the use of pass-gate transistors and also isolating the read current path by using transistors of minimum feature size; less leakage current and power than the 6T SRAM. Also, greater immunity to process variations was achieved in the 9T SRAM cell as well as the 6T SRAM. With less vulnerable to device mismatch because a symmetric approach for the 6T and 9T SRAM designs used. In addition, the 9T SRAM had extra three transistors for read assist, write assist, word line boosting schemes which helped to minimize any effect due to process variation.

## REFERENCES

[1]. Ezeogu, Apollos, "Process Variation Aware Non-Volatile (Memristive) 9T SRAM Memory Design in Nano-CMOS Technologies", M.Sc. Theses submitted to University of Bristol, United Kingdom, Oct. 2013.

*[2]. K. Dhanumjaya, MN. Giri Prasad, K. PAdmaraju, M. Raja Reddy, " Low Power and Improved Read Stability Cache Design in 45nm Technology", International Journal of Engineering Research and Development eISSN : 2278-067X, pISSN : 2278-800X, www.ijerd.com Volume 2, Issue 2 (July  2012), PP. 01-07*

[3]. G. K. Reddy, Kapil Jainwal, Jawar Singh and Saraju P. Mohanty, " Process Variation Tolerant 9T SRAM Bitcell Design", Quality Electronic Design (ISQED), 2012 13th International Symposium on 19-21 March 2012, pp 493 - 497, Santa Clara, CA.

[4]. S. Mutoh, T. Douseki, Y. Matsuya, T. Aoki, S. Shigematsu, and J. Yamada, "1-V power supply high-speed digital circuit    technology with multi threshold-voltage CMOS," IEEE J. Solid-State Circuits, Vol. 30, No. 8, pp. 847–854, 1995.

[5]. Govind Prasad, "Design of Low Power and High Stable Proposed SRAM cell Structure", International Journal of VLSI and Embedded Systems-IJVES ISSN: 2249 – 6556, 2013.

[6]. Evelyn Grossar, Michele Stucchi, Karen Maex, and Wim Dehaene, " Read Stability and Write-Ability Analysis of SRAM Cells for Nanometer Technologies", IEEE Journal of Solid-State Circuits, vol. 41, No.11, November 2006.

[7]. Archna bai, " SRAM Modelling for Read Stability and Write Ability Cell , International Journal of Emerging Technologies in Computational and Applied Sciences, 2 (1), Aug.-Nov., 2012, pp. 26-31.

[8]. Kaushik Roy, Saibal mukhopadhyay and Maymoodi - Meimand, "Leakage Current Mechanisms and Leakage Reduction Techniques in Deep-Sub micrometer CMOS Circuits", Proceedings of the IEEE, Vol. 91, No. 2, February , 2003.

[9]. Jawar Singh, Jimson Mathew, Saraju P. Mohanty and Dhiraj K. Pradhan, "Statistical Analysis of Steady State Leakage Currents in Nano-CMOS Devices", Published in: Norchip,19-20 Nov. 2007 pp. 1 - 4, Aalborg.

[10]. Benton H. Calhoun and Anantha Chandrakasan, "Analyzing Static Noise Margin for Subthreshold SRAM in 65nm CMOS" Proceedings of ESSCIRC, 2005.

*[11]. H. Kim and K. Roy, "Dynamic Vt SRAM: A leakage tolerant cache memory for low voltage microprocessors," presented at the Int. Symp. Low Power Electronics and Design, Monterey, CA, Aug. 2002.*

[12]. K. Seta, H. Hara, T. Kuroda, M. Kakumu, and T. Sakurai, "50% active power saving without speed degradation using stand-by power reduction (SPR) circuit," ISSCC Dig. Tech.Papers, pp. 318–319, 1995.

[13]. H. Kawaguchi, K. Nose, and T. Sakurai, "A CMOS scheme for 0.5V supply voltage with pico ampere standby current,"in Dig. Tech. Papers IEEE Int. Solid-State Circuits Conf., 1998, pp. 192–193.

[14]. K. Dhanumjay, M. Sudha, Dr.MN.Giri Prasad, K. Padmaraju "Cell stability Analysis of Conventional 6T Dynamic 8T SRAM Cell in 45nm Technology", "International Journal of VLSI design & Communication Systems (VLSICS), Vol. 3, No.2, pp. 41 - 51, April, 2012

[15]. Agarwal, H. Li, and K. Roy, "DRG-Cache: A data retention gated ground cache for low power," in Proceedings. 39th Design Automation Conference, 2002, pp. 473–478.

[16]. Mutyam M, Narayanan V. "Working with Process Variation Aware Cache". Design, Automation & Test in Europe Conference & Exhibition p1-6, 2007.